4

# MODEL OF ELECTROSTATIC ACTUATED DEFORMABLE MIRROR USING STRONGLY COUPLED ELECTRO-MECHANICAL FINITE ELEMENT

*Rochus V.[1], Golinval J.-C.[1], Louis C.[2], Mendez C.[2], Klapka I.[2]*

[1]University of Liège, LTAS, Vibrations et Identification des Structures,
Chemin des Chevreuils, 1, B52, 4000 Liège Belgique
[2] Open Engineering sa,
Rue des Chasseurs Ardennais, B-4031 Angleur, Belgique

## ABSTRACT

The aim of this paper is to deal with multi-physics simulation of micro-electro-mechanical systems (MEMS) based on an advanced numerical methodology. MEMS are very small devices in which electric as well as mechanical and fluid phenomena appear and interact. Because of their microscopic scale, strong coupling effects arise between the different physical fields, and some forces, which were negligible at macroscopic scale, have to be taken into account. In order to accurately design such micro-electro-mechanical systems, it is of primary importance to be able to handle the strong coupling between the electric and the mechanical fields. In this paper, the finite element method (FEM) is used to model the strong coupled electro-mechanical interactions and to perform static and transient analyses taking into account large mesh displacements. These analyses will be used to study the behaviour of electrostatically actuated micro-mirrors.

## 1. INTRODUCTION

Classical methods used to simulate the coupling between electric and mechanical fields are usually based on staggered procedures, which consist in computing quasi-static configurations using two separate models: a structural model loaded by electrostatic forces predicted at the current iteration step and an electrostatic model defined on the current deformed structure. Staggered iterations then lead to the static equilibrium position. In this paper, a fully coupled electro-mechanical FE formulation is proposed, which allows static equilibrium positions to be computed in a non-staggered way, and which provides fully consistent tangent stiffness matrices that can be used for transient analyses.

In the classical approach, staggered coupling procedures are typically used to simulate electro-mechanically coupled systems. The electrostatic and the mechanical domains are discretised and solved independently in different analysis steps. Iterations are then performed: the electrostatic field is first computed using, for instance, the boundary element method or the finite element method and it provides the electrostatic forces acting on the mechanical structure. Then a mechanical FEM code computes the structure deformation under the effect of electrostatic forces. The deformed structure defines new boundaries for the electrostatic problem and the electric field has to be computed again. This method is commonly presented in the literature (see e.g. Lee and al. [1]) and is illustrated in figure 1. The staggered method can also be understood as block-Gauss Seidel procedure.

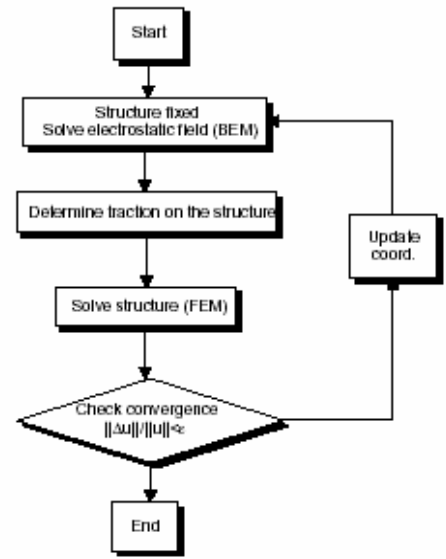

Figure 1: Staggered Method.

The method proposed here consists in considering the monolithically coupled problem. The electric and mechanical fields are computed *simultaneously* in a unified formulation. Since the problem is non-linear, it must be solved by an iterative algorithm such as the Newton-Raphson method or Riks-Crisfield method. The solution strategy consists in the following steps. Given an electric potential applied on the boundaries of the structure, a first solution is obtained by considering the coupled problem linearised around an initial configuration. The resulting structural deformation then defines a modified electric domain and a new linearised problem is defined around the modified configuration. This process continues until the solution has converged, namely until the electric and mechanical equilibrium are satisfied up to a prescribed tolerance (Figure 2).

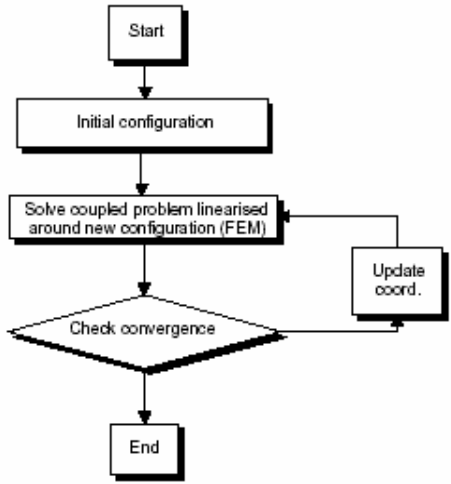

Figure 2: Monolithical Method.

The monolithic formulation has many advantages compared to the staggered method. Indeed the tangent stiffness matrix can be explicitly constructed for the entire system. Therefore, the stability and the natural frequency of the electrostatically coupled structure can easily be evaluated around a given equilibrium configuration [3]. Also the staggered coupled method becomes numerically unstable when the pull-in voltage is reached [2]. Finally, the coupled method permits to pass over the pull-in voltage.
Strong coupled methods are also proposed by ANSYS and FEMLAB to obtain the tangent stiffness matrix.

**1.1. ANSYS Coupled Formulation**

The software ANSYS proposes different solutions to achieve the equilibrium between electrostatic forces and mechanical structure [2,4,5]. The first one is a sequential coupling electro-mechanical solver called *ESSOLV* which uses the staggered methodology as presented previously. The second possibility is to use *TRANS126* element which is a strong coupled transducer formed by 2-nodes line [5]. The vacuum between the plates is discretised by one-dimensional condensers

To simulate very strong coupling in two dimensions *TRANS109* element is usually used [4]. This transducer is based on virtual principle. It is represented by a finite element 3-nodes triangle with electric and mechanical degrees of freedom. A standard Ritz-Galerkin approach is applied to compute the energy in the triangle. The forces in each node are obtained by deriving this energy by the nodes displacement. This methodology is used when the capacitance-stroke relation is difficult to obtain. The advantage of this element is its very accurate precision to compute the electrostatic forces and the electro-mechanical coupling, but the convergence is sensitive to the mesh discretisation and it is less robust for devices that experience large deformation [2]. An extension in three dimensions of this element has been performed by Avdeev [6]. It is called ``3D strongly coupled tetrahedral transducer''. Displacement and voltage are the variables assigned to the nodes. The energy in one element is computed by the relation:

$$W = \frac{1}{2}\int_\Omega \varepsilon_0 (\nabla V)^2 d\Omega = \frac{\varepsilon_0 \Omega}{2} \sum_{i=1}^{4} \sum_{j=1}^{4} \frac{\mathbf{n}_i . \mathbf{n}_j}{h_i h_j} V_i V_j$$

where $V$ is the electric potential, $\Omega$ the volume, $h_i$ is the tetrahedral altitudes and $\mathbf{n}_i$ the interface normal vector of the plane $i$ as shown in figure 3.

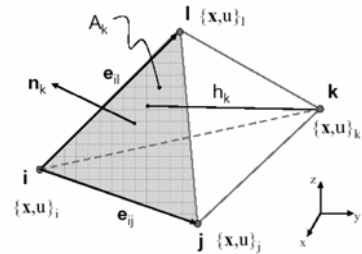

Figure 3: Avdeev Coupled element.

To compute electrostatic forces and tangent stiffness matrix, the altitudes and the normal vector are derived by the displacement of the nodes. This element as well as the *TRANS109* element, has an automatic mesh morphing based on equilibrium considerations. The initial mesh is automatically transformed to follow the displacement of the structure according to the force equilibrium criteria. With this method the electro-mechanical coupling is modelled strongly, but the formulation has been implemented only for linear tetrahedral elements.

**1.2. FEMLAB Coupled Formulation**

FEMLAB Software proposes a hybrid method to compute the electro-mechanical coupling. The mechanical structure is modelled with a classical mechanical model. For the electric domain new elements are created taking mesh deformation into account. The electrostatic field is solved and the electric forces are computed using the relation [7]:

$$\mathbf{F}_{elec} = \frac{\varepsilon \mathbf{E}^2}{2} = \frac{\varepsilon (\nabla V)^2}{2}$$

where **E** is the electric field.

The deformed electrostatic mesh is based on the coordinate transformation of the electric mesh. A new system to solve is added to the usual purely electric and mechanical problem creating a relation between the degrees of freedom of the mechanical structure and the position of the deformed electric mesh. The final system is the following:

$$\begin{cases} \mathbf{K}_{qq}\mathbf{q} = \mathbf{F}_{elec}(V) \\ \mathbf{K}_{uu}(\mathbf{u})\mathbf{u} = \mathbf{F}_{boundary}(V) \\ \mathbf{K}_{VV}(\mathbf{u},V)\mathbf{V} = \mathbf{V}_{boundary} \end{cases}$$

The first equation is the pure mechanical system taking the electrostatic forces $\mathbf{F}_{elec}$ into account. **q** is the displacements vector of the mechanical structure. The second system is the equilibrium equation of the mesh. **u** is the displacement of the mesh nodes. Finally, the third equation is the electrostatic system depending on the displacement of the mesh **u**. The electro-mechanical coupling appears through this deformed mesh. This methodology provides a strong electro-mechanical coupling. Problems appear in the electrostatic forces computation. First and second order elements do not provide the same converged electrostatic forces.

## 2. STRONG FINITE ELEMENT FORMULATION

The finite element formulation proposed in this paper is based on a variational principle applied to the entire energy of the problem. The energy includes together the electric and mechanical energies. The expression of the energy density of the system results from thermodynamical considerations [8]. Depending on the chosen variables, different thermodynamics functions have to be used. If the mechanical strain and the electric field are chosen, electric Gibbs' free energy density [9] or electric enthalpy [10] have to be considered. The difference between these two expressions appears in the thermal variables. Usually when analyzing piezoelectric behaviour of systems, the electric Gibbs' energy **G** is considered:

$$\mathbf{G} = \frac{1}{2}\mathbf{S}^T\mathbf{T} - \frac{1}{2}\mathbf{D}^T\mathbf{E}$$

where **T** is the stress tensor, **S** the strain tensor, **D** the electric displacement tensor and **E** the electric field.

The coupled electromechanical equations are obtained using a variational approach [11]. It leads to a consistent way to derive a finite element discretisation for the coupled electro-mechanical problem. Starting from the energy of the coupled system, nodal forces are obtained for an element by derivation of the energy. The tangent stiffness matrix of the coupled problem is then obtained by linearisation of the equilibrium equations in the vicinity of an equilibrium position.

The total energy on a volume Ω representing the electro-mechanical domain is:

$$W_{int} = \int_\Omega (\frac{1}{2}\mathbf{S}^T\mathbf{T} - \frac{1}{2}\mathbf{D}^T\mathbf{E})d\Omega = W_m - W_e$$

assuming linear elasticity and constant permittivity. $W_m$ is the mechanical energy and $W_e$ the electric one. The integration volume for the electric energy $W_e$ depends on the mechanical displacement **u**. For the mechanical energy $W_m$ however the integration domain is the reference (i.e. undeformed) domain since **u** represents the displacement as a function of the initial positions (Lagrangian description). The constitutive equations and the compatibility relations for both fields are used. The variation of the internal energy due to the virtual displacement (compatible with kinematical constraints) yields the mechanical internal forces and its variation due to virtual perturbations of the potential yields the electric charges:

$$\begin{cases} \mathbf{f}_m.\delta\mathbf{u} = \delta_u W_{int} = \delta_u W_m - \delta_u W_e \\ q_e.\delta\phi = \delta_\phi W_{int} = \delta_\phi W_m - \delta_\phi W_e \end{cases}$$

Three of these variations can be evaluated in a straightforward way. In fact, $\delta_u W_m$ and $\delta_\phi W_e$ can be treated as in the standard variational calculus for uncoupled electrostatics and mechanics. The mechanical energy is independent on the voltage: $\delta_\phi W_m = 0$.

On the other hand, the variation of the electric energy due to the displacement **u** is not common. To compare the energy before and after variation, they have to be expressed on the same reference domain. In order to show the dependence of the volume of the electric part with respect to the structural displacement, an initial volume Ω is considered and a perturbation of the displacement $\delta\mathbf{u}$ is applied to obtain the perturbed volume Ω.

After developments the variation of the electric energy corresponding to virtual displacements can also be written as

$$\mathbf{f}_{elec}.\delta\mathbf{u} = \frac{1}{2}\int_\Omega \mathbf{D}^T\mathbf{F}(\nabla\delta\mathbf{u})d\Omega$$

To obtain the tangent stiffness matrix the linearisation of the internal forces around an equilibrium position ($\mathbf{u}_e, \phi_e$) is performed and gives us:

$$\begin{cases} \mathbf{f}_m \approx \mathbf{f}_m^e + \left(\dfrac{\partial^2 W_m}{\partial u^2} - \dfrac{\partial^2 W_e}{\partial u^2}\right)d\mathbf{u} - \dfrac{\partial^2 W_e}{\partial \phi \partial u}d\phi \\ q_e \approx q_e^e - \dfrac{\partial^2 W_e}{\partial u \partial \phi}d\mathbf{u} - \dfrac{\partial^2 W_e}{\partial \phi^2}d\phi \end{cases}$$

which can be written in the matrix form:

$$\begin{pmatrix} \dfrac{\partial^2 W_m}{\partial u^2} - \dfrac{\partial^2 W_e}{\partial u^2} & \dfrac{\partial^2 W_e}{\partial u \partial \phi} \\ -\dfrac{\partial^2 W_e}{\partial u \partial \phi} & \dfrac{\partial^2 W_e}{\partial \phi^2} \end{pmatrix} \begin{pmatrix} du \\ d\phi \end{pmatrix} = \begin{pmatrix} \Delta f_m \\ \Delta q_e \end{pmatrix}$$

$$\begin{pmatrix} K_{uu} & K_{u\phi} \\ K_{\phi u} & K_{\phi\phi} \end{pmatrix} \begin{pmatrix} du \\ d\phi \end{pmatrix} = \begin{pmatrix} \Delta f_m \\ \Delta q_e \end{pmatrix}$$

The tangent stiffness matrix takes into account the electromechanical coupling in the system. It is observed that the coupling matrix is symmetric. The matrices $\dfrac{\partial^2 W_e}{\partial \phi^2}$ and $\dfrac{\partial^2 W_m}{\partial u^2}$ are the same as the stiffness matrices of the purely electric problem and the purely mechanical problem. Then only two new terms $\dfrac{\partial^2 W_e}{\partial u^2}$ and $\dfrac{\partial^2 W_e}{\partial u \partial \phi}$ have to be computed. After long developments an analytical expression of the tangent stiffness matrix is obtain using the following expressions:

$$\begin{cases} \delta_{u\phi} W_e = \int_\Omega (\nabla \delta \mathbf{u})^T \mathbf{F}_1 (\nabla \delta \phi) d\Omega \\ \delta_{uu} W_e = \int_\Omega (\nabla \delta \mathbf{u})^T \mathbf{F}_2 (\nabla \delta \mathbf{u}) d\Omega \end{cases}$$

where $\mathbf{F}_1$ and $\mathbf{F}_2$ are matrices depending on the electric field. Thanks to the tangent stiffness matrix a dynamical study may be performed on electro-mechanical problem.

## 3. APPLICATIONS

This strong finite element formulation has been implemented on the commercial code OOFELIE(r). Both following applications have been realised using it.

### 3.1. Clamped-Clamped Beam Micro-Mirror

The micro-device studied here is a micro-bridge who may be used as an electrostatically actuated micro-mirror. The dimensions of the beam are the following: a length of 300 μm, a thickness of 0.5μm and the gap between the electrodes is about 6μm. Due to the large aspect ratio, large displacement hypothesis has to be used. Three small electrodes are deposited on the substrate to actuate the beam. The length of these electrodes is about 60 μm and a voltage $V_1$, $V_2$ and $V_3$ may be applied between one electrode and the beam.

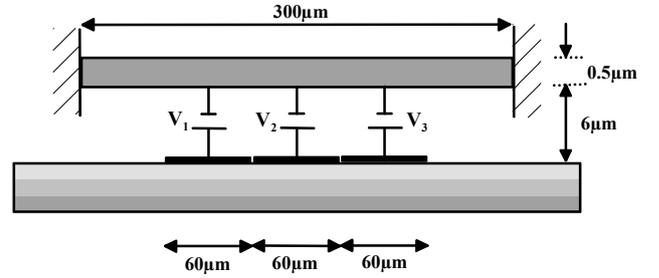

Figure 4 : Clamped-Clamped Beam Micro-Mirror.

Two different configurations will be considered: a voltage $V_2$ is applied to the centred electrode, then $V_1$ and $V_3$ will be set to a same predefined voltage. First a static study is performed to evaluate the static pull-in voltage. Then the dynamic behaviour of both configurations is studied when a voltage is suddenly applied between the electrodes.

*3.1.1. Static Pull-in Voltage*

When a voltage is applied between the membrane and the electrodes, electrostatic forces appear which force the membrane to bend. When the applied voltage increases, the electrostatic forces become dominant and for some limit value, the plates might stick together. The corresponding critical voltage is called the static pull-in voltage. It is one of the most important design parameters in this type of micro-systems. Indeed if the system achieved the pull-in state, the mirror may be destroyed.

Using a Riks-Crisfield algorithm the displacement of the centre of the beam may be computed when the voltage changes. The static pull-in voltage is estimated to 129.3 V for the centred electrode and to 141.6 V for the two electrodes as shown in figure 5. The displacement necessary to reach the static pull-in voltage is larger in the case of two electrodes than when there is only one.

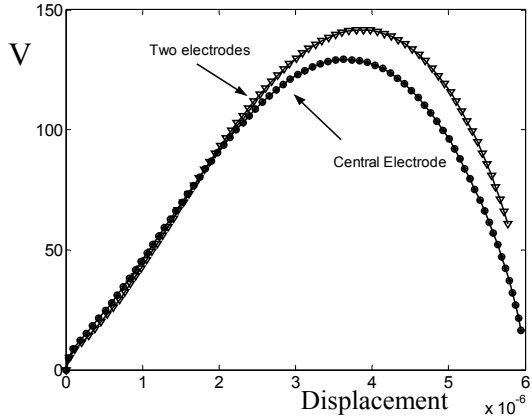

Figure 5: Static Pull-in Voltage.

*3.1.2. Dynamic Pull-in Voltage*

In this section, the transient dynamic response of the system is computed when the voltage is suddenly applied between the two electrodes i.e. when the time-history of the voltage corresponds to a step function as illustrated in figure 6.

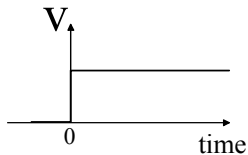

Figure 6: Step of voltage applied to the beam.

To investigate the dynamic pull-in, several voltages V were considered and applied to the beam as a step in time, the system being at rest initially. Using the Newmark scheme [12] to time integrate the problem one obtains the results depicted in the phase diagram in figure 7 and figure 8.

From the dynamic study of the electrostatically actuated beam behaviour, a new parameter is defined: the dynamic pull-in voltage. The definition of this voltage is the following:

*The dynamic pull-in voltage is defined as the voltage amplitude such that, when applied suddenly, it leads to the dynamical instability of the system.*

The dynamic pull-in voltage in the case of a single centred electrode is about 114.8 V, 10.5% lower than the static pull-in voltage with the same hypothesis. For the second problem, the dynamic pull-in voltage is about 125.1V, i.e. 11.6 % lower than the static one. So due to the dynamic effect the system may become unstable before reaching the static pull-in voltage.

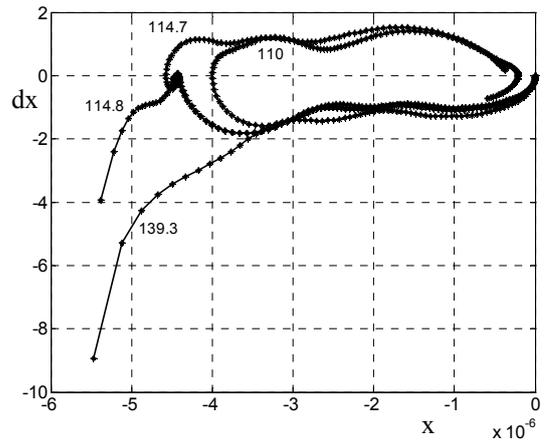

Figure 7: Phase diagram for the centred electrode.

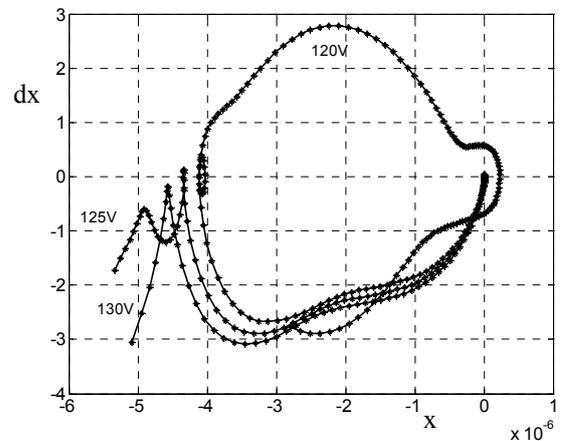

Figure 8 : Phase diagram for the two electrodes.

**3.2. Circular Micro-Mirror**

The second application example treated in this paper is the 37-channel Micromachined Deformable Mirror System [13]. This micro-mirror is fabricated by the company OKO technologies in cooperation with the Institute of Microelectronic and Submicron Technology of the University of Delft. It consists in a circular deformable membrane with a diameter of 15mm which is coated to form the mirror. 37 actuators are located under the membrane as shown in figure 9. The mirror shape is controlled by applying voltage between the membrane and some actuator electrodes. This micro-device is usually used for fast dynamic correction of low-order optical aberration such as defocus, astigmatism, coma,… in laser telescopes, ophthalmology, displays and general imaging optics. So it is important to know exactly the shape of the mirror during its actuation.

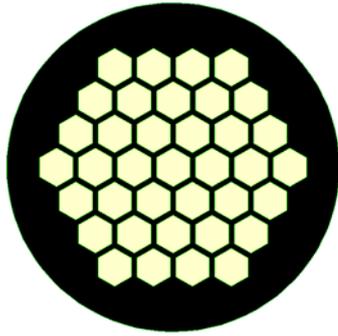

Figure 9: OKO technology micro-mirror.

A voltage of 100 V is applied to the electrodes. First one fourth of the electrodes are activated. The applied electric potential and the deformation are presented in figure 10. Then a ring of electrodes is activated and the results are shown in figure 11.

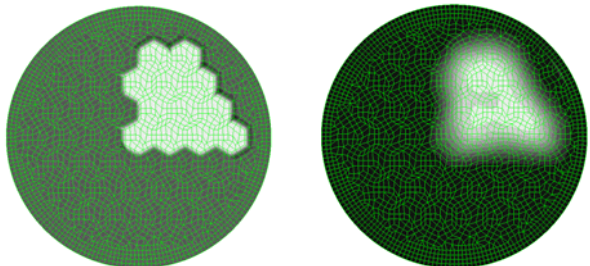

Figure 10: Dissymmetric Deformation.

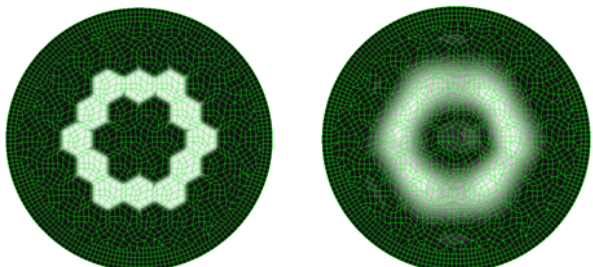

Figure 11: Circular deformation

## 4. CONCLUSIONS

The method proposed in this paper is a strong coupled formulation to compute the electro-mechanical coupling. This method differs from the other strong coupling methods due to its thermodynamics energy origin. The computation of electrostatic force derives from this energy and a second derivative provides the analytical expression of the tangent stiffness matrix. These expressions may be implemented for different types of elements (and not only for the tetrahedral linear element as in ANSYS). The electrostatic forces may be computed with linear and quadratic elements and converge to a single solution (not as FEMLAB). This methodology has been implemented in the commercial software called *oofelie* and has been used to study the dynamic behaviour of micro-mirrors.

## 5. ACKNOWLEGMENTS

The first author acknowledges the financial support of the Belgian National Fund for Scientific Research. Part of the work presented in this text is supported by the Communauté Française de Belgique – Direction Générale de la Recherche Scientifique in the framework Actions de Recherche Concertées (convention ARC 03/08-298).